# Continuous QoE Prediction Based on WaveNet

Phan Xuan Tan
SIT Research Laboratories
Shibaura Institute of Technology
Tokyo, Japan
tanpx@shibaura-it.ac.jp

Tho Nguyen Duc
Graduate School of Engineering and Science
Shibaura Institute of Technology
Tokyo, Japan
mg18502@shibaura-it.ac.jp

Chanh Minh Tran
Graduate School of Engineering and Science
Shibaura Institute of Technology
Tokyo, Japan
ma18502@shibaura-it.ac.jp

Eiji Kamioka
Graduate School of Engineering and Science
Shibaura Institute of Technology
Tokyo, Japan
kamioka@shibaura-it.ac.jp

## ABSTRACT

Continuous QoE prediction is crucial in the purpose of maximizing viewer satisfaction, by which video service providers could improve the revenue. Continuously predicting QoE is challenging since it requires QoE models that are capable of capturing the complex dependencies among QoE influence factors. The existing approaches that utilize Long-Short-Term-Memory (LSTM) network successfully model such long-term dependencies, providing the superior QoE prediction performance. However, the inherent drawback of sequential computing of LSTM will result in high computational cost in training and prediction tasks. Recently, WaveNet, a deep neural network for generating raw audio waveform, has been introduced. Immediately, it gains a great attention since it successfully leverages the characteristic of parallel computing of causal convolution and dilated convolution to deal with time-series data (e.g., audio signal). Being inspired by the success of WaveNet, in this paper, we propose WaveNet-based QoE model for continuous QoE prediction in video streaming services. The model is trained and tested upon on two publicly available databases, namely, LFOVIA Video QoE and LIVE Mobile Stall Video II. The experimental results demonstrate that the proposed model outperforms the baselines models in terms of processing time, while maintaining sufficient accuracy.

## 1. INTRODUCTION

Nowadays, online video has increasingly become the most dominant services on the Internet. According to recent study and forecast, the global video traffic will grow threefold between 2016 and 2021 [1]. Since small degradation in the perceived video quality can significantly influence the acceptance of service, video streaming services have high requirements on quality of experience (QoE). In order to increase the revenue, it is necessary for video service providers to mark QoE enhancement with high priority. The presence of Ultra High Definition (UHD) videos, 3D videos and the rapidly growing number of subscribers and high-resolution mobile devices cause the bandwidth starvation and unstable network condition, resulting in QoE deterioration. Therefore, it is important to continuously quantify the perceptual QoE of the streaming video users so that the QoE deterioration can be alleviated. However, the continuous prediction of QoE is challenging since it is determined by complex dynamic interactions among QoE influence factors. In such a situation, Long-Short-Term-Memory (LSTM)-based QoE model has been recently introduced [2]. This model is actually a network of cascaded LSTM blocks to capture the nonlinearities and the complex temporal dependencies involved in the time varying QoE. As a result, the model provides state-of-the-art QoE prediction performance. However, LSTM theoretically processes the data in sequential manner. It means that, the output is generated after the previous one. This significantly leads to the high computation cost in both training and prediction phases. Convolutional architecture, on the other hand, can provide potential benefit in terms of computing time due to its inherent characteristic of parallelization. WaveNet [3], a deep neural network has recently been introduced to grasp the strength of convolutional networks to generate wideband raw audio waveform which is recognized as time-series data. The model successfully deals with long-range temporal independencies of raw audio data, while performing the prediction in parallel. The success of WaveNet inspired us to consider such a convolution sequence modeling method in QoE prediction for the improvement of training time and prediction time, while maintaining longer effective memory.

In this paper, we propose WaveNet-QoE, a continuous QoE prediction model which takes advantages of parallel computing characteristic of convolutional networks, for better QoE prediction performance. The key contributions of the paper are briefly summarized in the following:

- A WaveNet-based QoE model is proposed for predicting continuous QoE based on WaveNet.

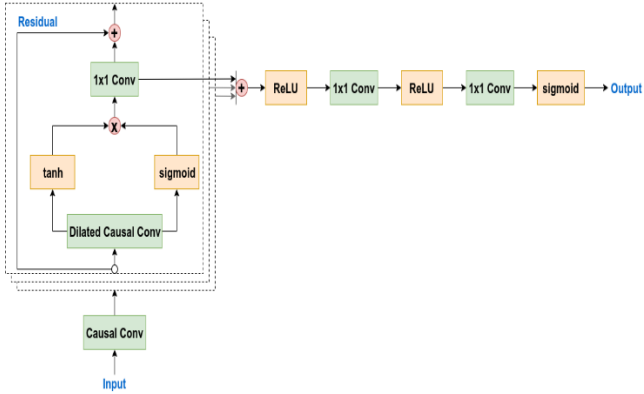

**Figure 1.** Overview of the entire architecture of WaveNet

- An evaluation of the proposed model is conducted on two continuous QoE databases, which demonstrates a competitive QoE prediction performance.

The rest of paper is organized as follows: section 2 provides the related work, whereas the proposal is described in section 3. The evaluation and discussion will be carried in section 4. Section 5 concludes the paper.

## 2. RELATED WORK

QoE modeling for video streaming services has recently received a lot of attentions from academic researchers due to its critical importance in QoE-aware applications. The challenges in QoE modeling are caused by the complex dependencies among QoE influence factors.

In literature, there exists numerous studies that address the challenge of continuously predicting QoE [4, 5, 6, 7, 8, 9, 10]. By considering perceptual video quality and rebuffering metrics, the authors in [8] proposed a nonlinear autoregressive with exogenous variables (NARX) model for continuous QoE prediction. This model was established based on LIVE-Netflix QoE Database [11] which comprises of many playout scenarios in which the presence of bitrate fluctuations and rebuffering events are taken into account. The authors in [10] proposed a QoE prediction model based on nonlinear state space (NLSS-QoE). Meanwhile, the time-varying QoE indexer to model nonlinearity and memory effects for predicting the continuous QoE was introduced in [9]. In fact, apart from the common QoE influence factors, human's memory also plays an important role in the assessment of the subjective QoE. Study in [11, 12] has proven the influence of primacy and recency effect when monitoring user's QoE. In short, QoE monitoring model should consider the long-term dependencies between events happening during the streaming session. This has pointed out the limitation of the above approaches.

Recently, huge research efforts have been carried out to utilize the Long Short-Term Memory (LSTM) approach to modeling and predicting several types of time-varying data. The work in [2] was one of the first studies to apply this method to QoE prediction in video streaming. The study proposed a continuous QoE prediction method utilizing the LSTM model (LSTM-QoE) to capture the nonlinearities and the complex temporal dependencies affecting the user's QoE. Although providing excellent QoE estimation performance, LSTM-QoE inherits the shortcoming of LSTM networks [13]. In fact, LSTM utilizes mainly sequential processing over time. Specifically, in the structure of LSTM, the sequential path exists from older past cells to the current cell, raising the question on its training and processing time. Recently, WaveNet [3], a novel deep learning model, has been introduced to leverage the convolutional architecture, reaching the state-of-the-art performance in generating raw audio waveform which is one of the variants of 1-D data. This leads to the potential success of convolutional sequence modeling. Therefore, in this paper, we propose WaveNet-QoE, a QoE prediction model based on WaveNet, to take the advantage of the convolutional architecture to boost up the training and processing time, while guaranteeing the competitive prediction accuracy.

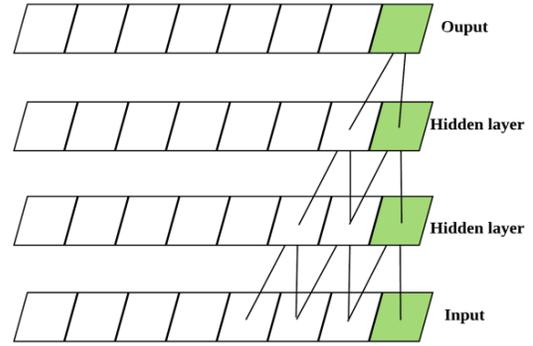

**Figure 2.** Causal Convolutional Layers

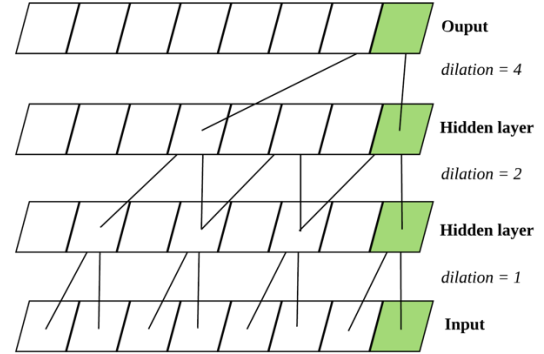

**Figure 3.** Dilated causal convolutional layers with dilation factor d = 2, the filter size k = 2, and number of layers L = 3.

## 3. PROPOSED METHOD

In this section, we first present the basics of WaveNet [3] in order to clarify the strength of this model in processing 1-D data. The WaveNet-based QoE prediction model is then described in detail.

### 3.1 Basics of WaveNet

Figure 1 depicts the overview of the residual block and the entire architecture of WaveNet [3]. It is a convolutional neural network used for directly estimating raw audio waveform. The joint probability of a waveform is factorized as a product of conditional probabilities as follows:

$$p(x) = \prod_{n=1}^{N} p(x_n|x_1, x_2, ..., x_{n-1}) \quad (1)$$

Each datapoint $x_n$ is therefore conditioned on the samples at all previous timesteps. The conditional probability distribution is modelled by a stack of convolutional layers. The core of WaveNet is causal convolutional architecture which is illustrated in Fig. 2. By using this, the model is guaranteed not to violate the order in which data is modelled: the prediction $p(x_n|x_1, x_2, ..., x_{n-1})$ emitted by the model at timestep $t$ cannot depend on any of the future timesteps $x_n, x_{n+1}, ..., x_N$. This idea is analogous as the masked convolution introduced in PixelCNNs model [14]. The model performs the predictions in sequential manner: after each sample is predicted, it is fed back into the network to predict the next sample. Due to the presence of convolutional architecture, the model does not have recurrent connections, resulting in faster training and predicting time.

In order to deal with the large filters (or high number of layers) of causal convolution, a remarkable characteristic is introduced in WaveNet. This characteristic is dilated causal convolution which is presented in Fig. 3. The purpose is to increase the receptive field by orders of magnitude, without increasing computational cost. In addition, with large receptive field, WaveNet can greatly capture the long-term dependencies in time-series data.

## 3.2 WaveNet-based QoE Model

QoE can be influenced by either technical (e.g., video codec, video buffer, video player) or perceptual factors (e.g., bitrate switching frequency, rebuffering frequency and duration) or both [15]. The aim of QoE modeling is to characterize the complex dynamic dependencies of those QoE influence factors, which is recognized as one of the most challenging tasks [8, 2].

Let $X_t$ be the vector of input features which are QoE influence factors. Thereby, the QoE prediction function can be formed as follows:

$$Y'_t = f(X_t, X_{t-1}, \ldots, X_{t-r+1}), r > 0 \quad (2)$$

where $Y'_t$ is instant QoE predicted at time $t$. Meanwhile $r$ stands for the lags of input which is defined as a fixed amount of passing time.

In the nature of the sequence modeling task, the causal constraint indicates that the prediction $Y'_t$ depends only on the inputs that have been previously observed. The goal of learning in QoE prediction is to find a network or nonlinear function $f$ that minimizes the expected loss between the subjective QoE and the predicted one. In this paper, the model is designed based on WaveNet whose architecture is shown in Fig. 1.

According to the original idea of WaveNet, in causal convolutional architecture (shown in Fig. 2), the predicted QoE at timestep $t$ will be provided upon the input ranging from $X_{t-r+1}$ to $X_t$. In this case, the lag value $r$ is considered as the size of receptive field. In other words, being different from LSTM architecture, WaveNet-QoE takes into account a sequence of input with the size of receptive field. Typically, the receptive field plays an important role in QoE modeling. The larger the receptive field is, the higher computational cost the model produces and vice versa. Therefore, it is necessary to determine the optimal value of receptive field for the proposed model. The receptive field is defined as follows [3]:

$$r = d^{L-1}k \quad (3)$$

where, $L$ is the number of layers, $d$ stands for dilated factor and $k$ denotes the filter size. It should be noted that Eq. (3) is only valid for a filter of size $k = 3$ and a dilated factor of value $d = 2$ [3]. In this paper, those factors, especially the receptive field size, will be estimated through the simple grid-search method. Determining the optimal range value of receptive field will be covered in future work.

## 4. EVALUATION

In this section, the performance of the proposed model is evaluated across two publicly available continuous QoE databases. Alternatively, the comparison among the proposed model and the baseline methods comprising of LSTM-QoE [2] and NLSS-QoE [10] is also conducted.

### 4.1 Input Features and Evaluation Database

*4.1.1 Feature Selection*
In order to provide the fair comparisons with baseline methods, the following four features are employed:

- Short Time Subjective Quality (STSQ) [4] refers to the perceptual video quality being rendered to the user.
- Playback Indicator (PI): a binary indicator specifying the current playback status: 1 for rebuffering and 0 for normal playback.
- Number of rebuffering events (NR): the number of interruption events happening from the beginning to current time instant of session
- Time elapsed since last rebuffering ($T_r$)

*4.1.2 Database Description*
In this paper, the following publicly continuous QoE databases are considered for the evaluation of the proposed model.

- LFOVIA Video QoE Database [12] consists of 36 distorted video sequences of 120 seconds duration. A training and test procedure is employed as described in [2, 10]. The databases are divided into different train-test sets. In each train-test set, there is only one video in the test set, whereas the training set includes the videos that do not have the same content and playout pattern as the test video. Thus, there are 36 train-test sets, and 25/36 videos are chosen for training the model for each test video.
- LIVE Mobile Stall Video Database II [16]: In this database, the distortion patterns are randomly distributed across videos. We first create 174 train-test sets corresponding to each of 174 videos in the database. For each train-test set, since the distortion patterns are randomly distributed across the videos, we then randomly choose 80% videos from the remaining 173 videos for training the model and perform evaluation over the test video.

### 4.2 Evaluation Criteria

The performance of the proposed WaveNet-QoE model is evaluated in terms of accuracy and computational cost.

To evaluate the accuracy, the following three metrics are considered: 1) Linear Correlation Coefficient (LCC), 2) Spearman Rank Order Correlation Coefficient (SROCC), and 3) Root Mean Squared Error (RMSE). The LCC and SROCC provide a quantification of the correlation between predicted QoE and subjective QoE in the database. Meanwhile, RMSE indicates the closeness between them.

To evaluate the computational cost, we consider the training time defined by s/epoch, which is recognized as the time to train an epoch and the testing time (ms).

### 4.3 Hyperparameter Selection

There are four network hyperparameters are considered in the model. They are:

- *Filter size* denoted by $k$
- *Number of filters* denoted by $n$.
- *Dilated factor* denoted by $d$
- *Receptive field* denoted by r

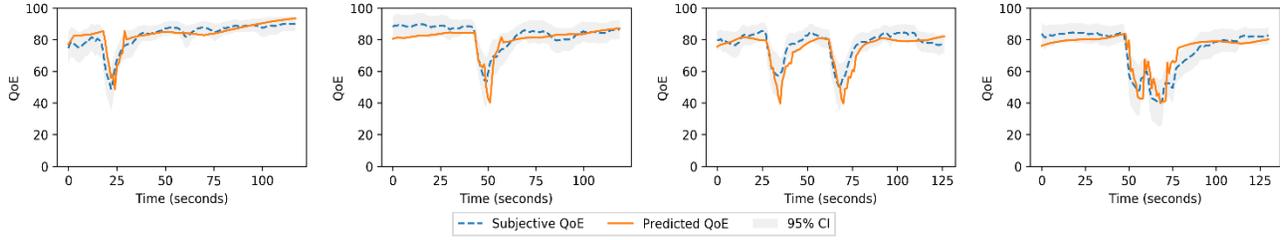

**Figure 4. QoE prediction performance of the proposed model in LFOVIA Video QoE Database.**

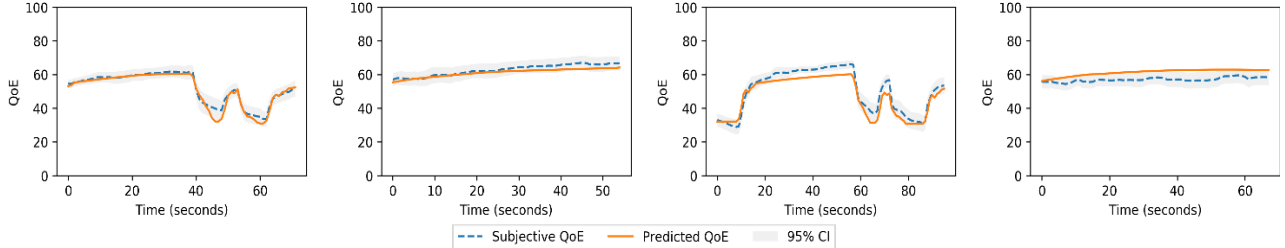

**Figure 5. QoE prediction performance of the proposed model in LIVE Mobile Stall Video Database II.**

Initially, the dilated factor is set to $d = 2$ which is the same as in [3] for the simplicity. Based on the Eq. 3, we then conducted a simple grid-search of the hyperparameter values to train the model on the training dataset, then evaluated its performance on the testing dataset. Table 1 tabulates the selected hyperparameters of our proposed model, whereas table 2 presents the optimizer algorithm and learning rate.

**Table 1. Hyperparameter for the best performance model**

| Architecture Hyperparameters | Derived values |
|---|---|
| Filter Size | $k = 2$ |
| Number of filters | $n = 32$ |
| Dilated factor | $d = 2$ |
| Receptive field | $r = 8$ |

**Table 2. Training hyperparameters**

| Learning rate | 0.001 |
|---|---|
| Optimizer algorithm | $Adam$ |

### 4.4 WaveNet-QoE Evaluation

In all our evaluations, we employ the best configuration as determined in Table 1 and 2 for the proposed model. During training, the data is fed to the network through an input layer with appropriate timesteps of 8. The sample rate is set to 1 second, and hence, while testing, the QoE is predicted with a granularity of 1 second. In other words, it is performed at every timestep.

Figures 4 and 5 illustrate the QoE prediction along with subjective QoE on the considered databases using the proposed WaveNet-QoE model. Accordingly, across all the scenarios, it may be observed that the proposed model does not overfit to the existing database, but instead attempts to accurately predict the varying trends in each dynamic QoE prediction. Despite maintaining a strong monotonic relationship with the ground truth dynamic QoE, the QoE predictions occasionally fall outside of 95% confidence interval. More specifically, when the rebuffering events frequently occur (as the fourth scenario in Fig. 4) during a streaming session, the model seems to underperform. The reason might be the absence of the other types of QoE influence factors (e.g., memory effect, video content) in the proposed model. This is understandable since the proposed model only produces the instantaneous prediction, whereas mathematical expressions are needed to involve such factors.

**Table 3. QoE prediction performance of the proposed model over the LFOVIA Video QoE Database.**

| QoE model | PCC | SROCC | RMSE |
|---|---|---|---|
| WaveNet-QoE | 0.790 | **0.888** | **6.757** |
| LSTM-QoE [2] | **0.800** | 0.730 | 9.560 |
| NLSS-QoE [10] | 0.767 | 0.685 | 7.590 |

**Table 4. QoE prediction performance of the proposed model over the LIVE Mobile Stall Video Database II.**

| QoE model | PCC | SROCC | RMSE |
|---|---|---|---|
| WaveNet-QoE | 0.792 | 0.851 | **6.665** |
| LSTM-QoE [2] | **0.878** | **0.862** | 7.080 |
| NLSS-QoE [10] | 0.680 | 0.590 | 9.520 |

**Table 5. Model training and predicting efficiency.**

| QoE model | Training time (s/epoch) | Inference time (ms) |
|---|---|---|
| WaveNet-QoE | **0.083** | **1.149** |
| LSTM-QoE [2] | 4.351 | 1.996 |

Tables 3, 4 present the comparison results in terms of accuracy between the proposed model and baseline models. Meanwhile, Table 5 provides the results of training and testing time. In general, the proposed model provides a competitive prediction performance. It can be observed that the proposed model outperforms NLSS-QoE in terms of PCC, SROCC and RMSE on both considered databases. In comparison with LSTM-QoE, even though presenting better RMSE value, the proposed model achieves relatively equivalent performance in terms of correlation between predicted QoE and the ground truth QoE, which is defined by PCC and SROCC. This can be explained by recalling the receptive field size of LSTM-QoE and the proposed model. While LSTM performs the prediction based

on all the past timesteps of data, the WaveNet-QoE, according to Eq. 2, considers a specific range of timesteps data, fitting in the size of receptive field. However, by leveraging the causal convolutional and dilated causal convolutional architecture, the proposed model yields superior computing time. According to Table 5, while LSTM-QoE takes about 4.351s to finish training an epoch, the proposed model is about 52 times faster. For prediction time, the proposed model spends only 1.149ms for each prediction, whereas, it takes about 1.996ms for LSTM-QoE for performing the same task.

## 5. Conclusion

This paper presents WaveNet-QoE, a deep continuous QoE prediction model that leverages the strength of convolutional architecture to achieve the competitive QoE performance. The model successfully combines causal filters and dilated convolutions to allow a larger receptive field, which is important to model long-range temporal dependencies in QoE data. In comparison with baseline methods which are built upon LSTM networks, the proposed model provides an extremely small training time and quick and high accurate prediction. The results demonstrate that the proposed model could be promisingly applied in any real-time QoE-aware application. Additionally, the high performance of WaveNet-based QoE prediction model indicates the potential of convolution sequence modeling.

# AUTHORS' BACKGROUND

*This form helps us to understand your paper better, the form itself will not be published. So please fill in every author's information.

*Position can be chosen from: master student, PhD candidate, assistant professor, lecturer, senior lecture, associate professor, full professor

| Your Name | Position | Email | Research Field | Personal website |
|---|---|---|---|---|
| Phan Xuan Tan | Assistant Professor | tanpx@shibaura-it.ac.jp | Image & Video processing, Multimedia technology, Networking | |
| Tho Nguyen Duc | Master student | mg18502@shibaura-it.ac.jp | Information & Communication System | |
| Chanh Tran Minh | Master student | ma18502@shibaura-it.ac.jp | Information & Communication System | |
| Eiji Kamioka | Professor | kamioka@shibaura-it.ac.jp | Information & Communication System | http://kamioka.net/index-e.html |